\documentclass[prl,twocolumn,superscriptaddress,msmath,amssymb]{revtex4}

\usepackage{psfig,bm}
\begin{document}
\psfigurepath{./figures}

\title{Partial spin freezing in the quasi-two-dimensional La$_{2}$(Cu,{\it Li})O$_4$}

\author{Y. Chen}
\affiliation{Condensed Matter and Thermal Physics, Los Alamos National Laboratory, Los Alamos, NM 87545}
\affiliation{NIST Center for Neutron Research, National Institute of Standards 
and Technology, Gaithersburg, MD 20899}
\affiliation{Dept.\ of Materials Science and Engineering, University of
Maryland, College Park, MD 20742}
\author{Wei Bao}
\affiliation{Condensed Matter and Thermal Physics, Los Alamos National Laboratory, Los Alamos, NM 87545}
\author{Y. Qiu}
\affiliation{NIST Center for Neutron Research, National Institute of Standards 
and Technology, Gaithersburg, MD 20899}
\affiliation{Dept.\ of Materials Science and Engineering, University of
Maryland, College Park, MD 20742}
\author{J.E. Lorenzo}
\affiliation{CNRS, BP 166X, F-38043, Grenoble, France}
\author{J.L. Sarrao}
\affiliation{Condensed Matter and Thermal Physics, Los Alamos National Laboratory, Los Alamos, NM 87545}
\author{D.L. Ho}
\affiliation{NIST Center for Neutron Research, National Institute of Standards 
and Technology, Gaithersburg, MD 20899}
\affiliation{Dept.\ of Materials Science and Engineering, University of
Maryland, College Park, MD 20742}
\author{Min Y. Lin}
\affiliation{NIST Center for Neutron Research, National Institute of Standards 
and Technology, Gaithersburg, MD 20899}
\affiliation{ExxonMobil Research and Engineering Company, Annandale, NJ 08801}

\date{\today}

\begin{abstract}
In conventional spin glasses, the magnetic interaction is not 
strongly anisotropic and the entire spin system freezes at low temperature. 
In La$_2$Cu$_{0.94}$Li$_{0.06}$O$_4$,
for which the in-plane exchange interaction dominates the interplane one, 
only a fraction of spins with antiferromagnetic correlations extending to 
neighboring planes become spin-glass. The remaining spins with only in-plane 
antiferromagnetic correlations remain spin-liquid at low temperature. 
Such a novel partial spin freezing out of a 
spin-liquid observed in this cold neutron scattering study 
is likely due to a delicate
balance between disorder and quantum fluctuations
in the quasi-two dimensional $S$=1/2 Heisenberg system.
\end{abstract}


\maketitle

The parent compound for high transition-temperature superconductors,
La$_2$CuO$_4$, is an antiferromagnetic (AF) insulator. Magnetic exchange
interaction $J$ between the nearest neighbor $S$=1/2 spins of Cu$^{2+}$ ions
in the CuO$_2$ plane is several orders of magnitude stronger than the
interplane exchange interaction, making quantum spin fluctuations an
essential ingredient for magnetic properties in the quasi-two-dimensional 
(2D) Heisenberg system\cite{la2dv2,la2smhao,2dheis}. 
The N\'{e}el temperature $T_N$ of La$_2$CuO$_4$ is suppressed rapidly 
by 2$-$3\% hole dopants such as Sr, Ba or Li\cite{nagano,Li214phs}, 
while it is suppressed with isovalent Cu
substitution at a much higher concentration close to the
site dilution percolating threshold\cite{ZnSr214}. 
The strong effect of holes has been shown to be related to induced
magnetic vortices, which are topological defects in
2D systems\cite{h_haas}. The paramagnetic phase exposed by hole doping
at $T \ll J/k_B$ is dominated by zero-point quantum spin 
fluctuations and is referred to as a quantum spin liquid\cite{2dheis}.
The approach of the N\'{e}el transition toward zero temperature 
promises detailed predictions for spin dynamics by extending the 
celebrated theory of critical phenomena for classical statistical 
systems to quantum statistical systems\cite{ja_hertz,2dheis}.

However, in a wide doping range of La$_{2}$Cu$_{1-x}$Li$_x$O$_4$
below $\sim$10~K, 
a spin-glass transition has 
been reported in muon spin rotation ($\mu$SR)\cite{Li214phs}, 
nuclear quadrupole resonance (NQR)\cite{Li214NQR} and 
magnetization\cite{Li214phs2} studies.
A similar magnetic phase diagram has also been reported for
La$_{2-x}$Sr$_x$CuO$_4$ and Y$_{1-x}$Ca$_x$Ba$_2$Cu$_3$O$_6$\cite{sg_drh,bjbquasi,fchou}.
In conventional spin glasses, magnetic interactions are more or
less isotropic in space, and the whole spin system is believed 
to be frozen in the spin-glass phase\cite{byoung}. 
If the spin-glass phase in hole-doped cuprates behaved as in 
conventional spin-glasses, the ground state would be a spin-glass instead of 
N\'{e}el order or quantum spin liquid, and as pointed out
by Hasselmann et~al.\cite{sg_cn}, the quantum critical point of
the AF phase would be preempted.

Recently, 2D spin fluctuations in La$_{2}$Cu$_{1-x}$Li$_x$O$_4$ 
($0.04\leq x\leq 0.1$) were observed to remain liquid-like below  
the spin-glass transition temperature $T_g$\cite{bao02c,bao04a}. 
The characteristic energy of 2D spin fluctuations 
saturates at a finite value below $\sim$50~K as expected for a 
quantum spin liquid\cite{2dheis}, 
instead of becoming zero 
at $T_g$ as for spin-glass materials\cite{byoung,FeAl_msm}.
To reconcile these apparently contradicting experimental results, 
we have conducted a thorough magnetic neutron scattering investigation 
of La$_2$Cu$_{0.94}$Li$_{0.06}$O$_4$ to search for spin-glass
behavior. We found that in addition to the 
liquid-like 2D dynamic spin correlations, the rest of spins which participate 
in almost 3D and quasi-3D correlations become frozen in the 
spin-glass transition. This partial spin freezing in the laminar
cuprate is distinctly different from total spin freezing
in conventional 3D spin-glass materials.
The observed phase separation into spin glass and spin liquid components 
of different dimensionality sheds light on a long-standing confusion 
surrounding the magnetic ground state in hole-doped cuprates.

The same single crystal sample of La$_2$Cu$_{0.94}$Li$_{0.06}$O$_4$
used previously\cite{bao02c} was investigated in this work. 
$T_g$$\approx$8~K was determined using $\mu$SR\cite{Li214phs}. 
The lattice parameters of the orthorhombic $Cmca$ unit cell are
$a$=5.332$\AA$, $b$=13.12$\AA$ and $c$=5.402$\AA$ at 15~K. Wave-vector
transfers {\bf q} near (000) and (100) in both the ($h0l$) and ($hk0$) 
reciprocal planes were investigated at NIST using the 30 meter high 
resolution small angle neutron scattering (SANS) instrument at NG7, 
and cold neutron triple-axis spectrometer SPINS.
We set the array detector of NG7-SANS to  
1 and 9 m, corresponding to a $q$ range from 0.012 to 0.39~$\AA^{-1}$ 
and from 0.0033 to 0.050~$\AA^{-1}$, respectively.  
At SPINS, the (002) reflection of pyrolytic 
graphite was used for both the monochromator and analyzer. Horizontal 
Soller slits of 80$^{\prime}$ were placed before and after the sample.
A cold BeO or Be filter was put before the analyzer to eliminate 
higher order neutron in the fixed $E_f$=3.7 or 5 meV configuration,
respectively.
Sample temperature was controlled by a pumped $^4$He cryostat which could
reach down to 1.5~K.

Hole induced ferromagnetic exchange has been theoretically proposed
in the CuO$_2$ plane\cite{sg_rjg,sg_cn}. 
Although long-range ferromagnetic order has never been observed,
there is the possibility of short-range ferromagnetic spin clusters 
which freeze in the spin-glass state. A powerful tool to probe such 
clusters is SANS\cite{FeAl_msm}.
\begin{figure}[t]
\vskip -3ex
 \centerline{
\psfig{file=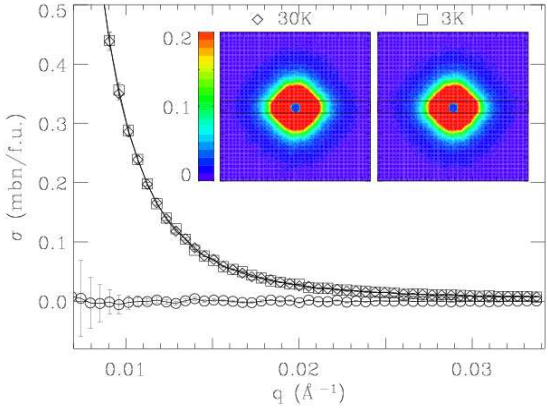,width=.95\columnwidth,angle=90,clip=}}
\vskip -3ex 
\caption{ \label{fig1} 
Measured SANS cross-section in the 10 pixels
wide rectangle shown in the inset at 3~K (squares) and 30~K
(diamonds), and the difference of the intensity between 3 K and 30 K
(circles) as a function of wave vector transfer $q$. Lines
are guide to the eye. Inset: the SANS pattern at 30 and 3~K.  
}
\end{figure}
Two reciprocal zones were studied, with incident 
beam parallel to the (001) or (010) direction, to cover
all spin orientations. The
experiments were carried out at 3, 10, 15, 30 and 80~K.  A
collection time of 1 or 2 hours per temperature provides good
statistics. No temperature dependence in the scattering patterns 
could be detected. The inset to Fig.~\ref{fig1} shows SANS patterns at
3 K and 30 K with incident beam parallel to the (001) direction.
Intensity at 3 and 30~K in the rectangular box on the 
SANS pattern is shown in the main frame.
The difference intensity (circles) fluctuates around zero,
and its standard deviation sets a upper limit of 
1.5$\times$10$^{-7}$ bn or
1.4$\times$10$^{-3} \mu_B$ per Cu for ferromagnetic moments 
in the clusters.  

\begin{figure}[t]
\vskip -3ex 
\centerline{
\psfig{file=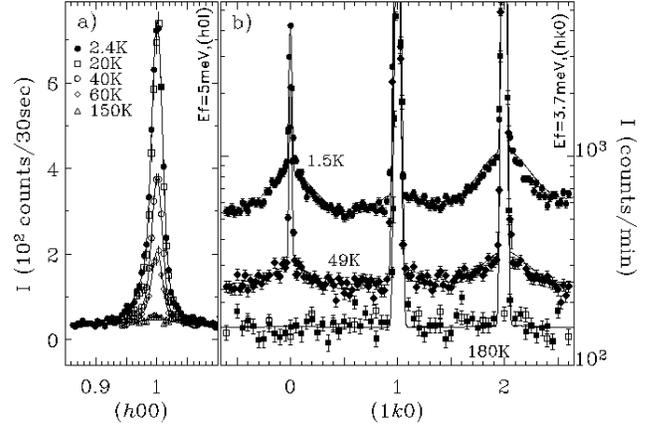,width=1.\columnwidth,angle=90,clip=}}
\vskip -2ex 
\caption{ \label{fig2} 
Magnetic quasielastic and elastic scattering along a) an
in-plane direction and b) the interlayer direction at various
temperatures. Open squares in b) was measured at (1.06,$k$,0) 
and represent background.
The solid lines are resolution convoluted 
$S^{3D}({\bf q},E)+S^{q3D}({\bf q},E)$ in Eq.~(\ref{eq1})-(\ref{eq2}).
}
\end{figure}

While no appreciable ferromagnetic signal was detected for
La$_2$Cu$_{0.94}$Li$_{0.06}$O$_4$, antiferromagnetic
scattering was observed along the rods 
perpendicular to the CuO$_2$ plane and intercepting the 
plane at the commensurate ($\pi,\pi$)-type Bragg points 
of the square lattice. This means that antiferromagnetic
correlations in the CuO$_2$ plane is chessboard-like,
which is simpler than in La$_{2-x}$Sr$_x$CuO$_4$\cite{Waki_Sr3}.
Elastic and quasielastic scans
through such a rod at (100) are shown in Fig.~\ref{fig2}(a)
at various temperatures. 
Inelastic scans have been presented
in \cite{bao02c}. There is little change in the peak width
in these scans, consistent with previous results of 
temperature independent in-plane correlation length 
for La$_2$Cu$_{0.95}$Li$_{0.05}$O$_4$\cite{la2dv2}
and La$_{2-x}$Sr$_x$CuO$_4$ (0.02$\leq x\leq$0.04)\cite{la2bkb}
below 300~K. Modeling the width of the rod in 
La$_2$Cu$_{0.94}$Li$_{0.06}$O$_4$ with Lorentian
\begin{equation}
\mathcal{L}^{\xi}(q)=\xi/\pi[1+(q\xi)^2],
 \label{eq0}
\end{equation}
$\xi_{\Box}^{3D}\ge 274 \AA$ from scan at (100) and $E=0$, 
$\xi_{\Box}^{q3D}\ge 70 \AA$ at (1,0.4,0) and $E=0$,
and $\xi_{\Box}^{2D}\ge 55\AA$ at finite $E$ from deconvolution. 
The $\Box$ denotes the in-plane component.
These planar AF clusters correlate in three different 
ways in the interlayer direction, giving rise to
almost 3D, quasi-3D and 2D magnetic correlations. 
Let us now examine the three components.

\begin{figure}[t]
\vskip -5ex
\centerline{
\psfig{file=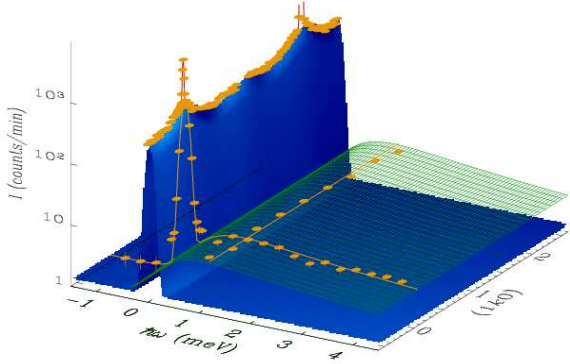,width=1.\columnwidth,angle=90,clip=}} 
\vskip -4ex
\caption{ \label{fig3} 
Measured $S({\bf q},E)$ as a function of $E$ and interlayer $k$ at 1.5~K, 
showing three color-coded magnetic components in Eq.~(5). 
The $E=0$ peak is energy-resolution limited. 
}
\end{figure}

Fig.~\ref{fig2}(b) shows elastic and quasielastic scans
along the rod in the interlayer direction at 1.5, 49 and 180~K. 
Magnetic intensity is composed of broad and sharp
peaks at magnetic Bragg points (100) and 
(120) of the parent compound. 
The (110) peak is temperature-independent thus nonmagnetic.
Fitting the broad
peaks to Eq.~(\ref{eq0}), we obtained an interlayer correlation 
length $\xi^{q3D}= 6.2(4)$~$\AA$. Again, no temperature dependence
can be detected for $\xi^{q3D}$ below 49~K. Thus, the quasi-3D 
spin correlations are typically three planes thick. For
the sharp peak, $\xi^{3D}\ge 168\AA$. 
Therefore, the number of correlated AF planes is more 
than 50, resembling 3D AF order.
Both the broad and sharp peaks are energy-resolution-limited
with the half-width-at-half-maximum $\le$~0.07 meV, 
as it is displayed in Fig.~\ref{fig3} and in \cite{bao02c}. 
However, they should not be regarded automatically
as from {\em static} magnetic order. Static magnetic signal 
was observed in $\mu$SR study\cite{Li214phs} only below $T_g$=8~K 
at the spin glass transition. Thus, the energy spectra
for the quasi-3D and almost 3D  
correlations may be modeled by $\mathcal{L}^{1/\epsilon}(E)$, with 
$\epsilon/h$ much smaller than 17 GHz=0.07 meV/$h$, the frequency 
resolution at SPINS, and
larger than the zero-field $\mu$SR static cutoff frequency of 
about 1 MHz\cite{Li214phs,bjbquasi} for $T> 8$~K.

The 2D AF correlations have been investigated 
in detail in \cite{bao02c}. The dynamic magnetic structure factor,
\begin{equation}
S^{2D}({\bf q},E)=\sum_{\bm{\tau}} 
\mathcal{L}^{\xi_{\Box}^{2D}}(\bm{\kappa}_{\Box})
\frac{\chi''(E)}{\pi \left(1-e^{-\hbar\omega/k_BT}\right)},
\label{eq4}
\end{equation}
has been measured for $E \le 4.2$~meV between 1.5 and 
150~K near a magnetic Bragg wave-vector ${\bm{\tau}}$,
and $\bm{\kappa}\equiv {\bf q}-{\bm{\tau}}$.
The almost 3D and quasi-3D spin correlations can be described,
respectively, by 
\begin{equation}
S^{3D}({\bf q},E)=I^{3D} \sum_{\bm{\tau}} 
\mathcal{L}^{\xi_{\Box}^{3D}}(\bm{\kappa}_{\Box})
\mathcal{L}^{\xi_{3D}}(k-\tau_k) \mathcal{L}^{1/\epsilon}(E) \label{eq1}
\end{equation}
and
\begin{equation}
S^{q3D}({\bf q},E)=I^{q3D} \sum_{\bm{\tau}} 
\mathcal{L}^{\xi_{\Box}^{q3D}}(\bm{\kappa}_{\Box}) 
\mathcal{L}^{\xi_{q3D}}(k-\tau_k) \mathcal{L}^{1/\epsilon}(E). \label{eq2}
\end{equation}
With negligible ferromagnetic correlations, the total
dynamic structure factor is a summation of Eq.~(\ref{eq4})-(\ref{eq2}),
\begin{equation}
S({\bf q},E)=S^{2D}({\bf q},E)+S^{q3D}({\bf q},E)+S^{3D}({\bf q},E).
\end{equation}
Of the four variables of $S({\bf q},E)$, {\bf q}$_{\Box}$ is fixed
at the ($\pi,\pi$)-type Bragg points.
To comprehend the composition of $S({\bf q},E)$, it is 
sufficient to plot $S({\bf q},E)$ as a function of $E$ and the 
interlayer wavenumber $k$. Such a plot of measured $S({\bf q},E)$
at 1.5~K is shown in Fig.~\ref{fig3}, with the
temperature and {\bf q} independent incoherent scattering 
at $E=0$ subtracted.
The sharp peak fitted by the red curve is from $S^{3D}({\bf q},E)$,
the narrow blue ridge at $E$=0 from $S^{q3D}({\bf q},E)$, and
the green surface from $S^{2D}({\bf q},E)$. The red peak at (100)
is about one/three orders of magnitude stronger than the peak intensity
of the blue/green surface, respectively.

\begin{figure}[t]
\centerline{
\psfig{file=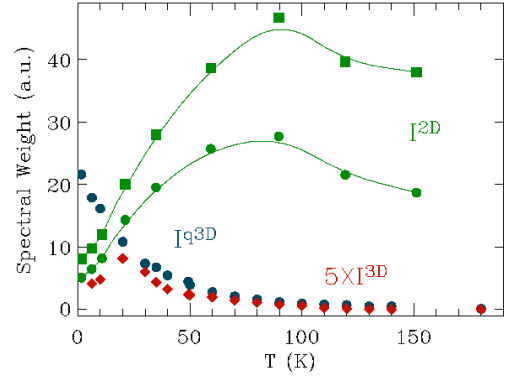,width=.8\columnwidth,angle=90,clip=}} 
\vskip -2ex
\caption{ \label{fig4} 
Temperature dependence of spectral weights
$I^{2D}$ (green), $I^{q3D}$ (blue) and $I^{3D}$ (red). 
See text for details.
}
\end{figure}

The spectral weights 
$\int d{\bf q} dE\, S^{3D}({\bf q},E)$$=$$I^{3D}$ and 
$\int d{\bf q} dE\, S^{q3D}({\bf q},E)$$=$$I^{q3D}$ can be obtained
by fitting resolution-convoluted Eq.~(\ref{eq1})-(\ref{eq2})
to scans in Fig.~\ref{fig2}(b). They are shown as a function of 
temperature in Fig.~\ref{fig4} with $I^{3D}$ magnified by a 
factor of 5 for clarity. For the 2D component, 
\begin{equation}
I^{2D}=\int dE\, \frac{2\chi''(E)}{\pi \left(1-e^{-\hbar\omega/k_BT}\right)}.
 \label{eq6}
\end{equation}
Green circles in Fig.~\ref{fig4} represent $I^{2D}$ with energy 
integration from $-0.5$ to 4.2 meV, the same interval as in 
experiment\cite{bao02c}. Green squares are obtained with a wider 
integration range, $|E| \le 10$~meV, using the analytical expression 
of $\chi''(E)$ in \cite{bao02c} to extrapolate to $E$=10 meV, 
where spin fluctuations were observed in La$_2$Cu$_{0.9}$Li$_{0.1}$O$_4$
using a thermal neutron spectrometer\cite{bao99a}.
Thus, the green symbols serve as the lower bound of
the 2D spectral weight which has $\pm \infty$ as the
integration limits. 

$I^{3D}$ and $I^{q3D}$ appear simultaneously below $\sim$150~K.
Their concave shape in Fig.~\ref{fig4} differ drastically 
from the usual convex-shape of a squared order parameter,
which was observed in $\mu$SR study below $T_g$=8~K\cite{Li214phs}. 
They are typical neutron scattering signal from  
slow {\em dynamic} spin correlations in spin-glasses\cite{FeAl_msm,sg_mh},
which fluctuate in the frequency window between 1 MHz 
and 17 GHz for $T> 8$~K, and below 1 MHz for $T< 8$~K.
Previously, energy-resolution-limited
neutron scattering from  
La$_{1.94}$Sr$_{0.06}$CuO$_4$ was observed to
have a similar temperature dependence as $I^{3D}$ in 
Fig.~\ref{fig4} and was attributed to
spin freezing\cite{bjbquasi}.
The kink at 20~K of $I^{3D}$ reflects an increased $T_g$ from 8~K 
to 20~K when probing frequency is increased from 1 MHz 
to 17 GHz\cite{bjbquasi,byoung}. 

The fact that $I^{3D}$ decreases below $T_g$(17MHz)$\approx$20~K 
while $I^{q3D}$ continues to increase indicates that 
the ``Edwards-Anderson order parameter''\cite{byoung,FeAl_msm,sg_mh} 
distributes only 
along the (1$k$0) line. In conventional spin-glasses, 
the ``Edwards-Anderson order parameter'' is more
isotropically distributed in the {\bf q}-space\cite{byoung,sg_mh,FeAl_msm}.
Thus, the spin-glass state in La$_2$Cu$_{0.94}$Li$_{0.06}$O$_4$
is characterized by interlayer disorder which upsets phase correlation
between large AF clusters in different CuO$_2$ planes. 
Another important difference from conventional spin-glasses in which 
all spins freeze below $T_g$ is that only a fraction of spins
freeze in La$_2$Cu$_{0.94}$Li$_{0.06}$O$_4$. 
Other spins in 2D correlations 
remain fluctuating down to 1.5~K. This is consistent with 
previous theory that quantum fluctuations prevent spin-glass
transition for 2D $S$=1/2 Heisenberg system\cite{bhatt}.
The spin-glass component in our sample has to acquire interlayer
correlations to achieve a higher dimension in order to be realized.
It appears that the lower critical dimension for a $S$=1/2 Heisenberg
quantum spin glass is between 2 and 3.
A further difference from conventional spin-glasses, for 
which one can measure the narrowing of magnetic 
spectrum toward $E$=0\cite{FeAl_msm}, is that
when $S^{3D}({\bf q},E)$ and $S^{q3D}({\bf q},E)$ in 
La$_2$Cu$_{0.94}$Li$_{0.06}$O$_4$ become detectable at about 150~K,
they are already energy-resolution-limited.
This property of $S^{3D}({\bf q},E)$ and $S^{q3D}({\bf q},E)$ resembles
the classic central peak phenomenon in the soft phonon
transition\cite{SrTiO}.
The disparate dynamics of the central peak and phonon are explained by
Halperin and Varma\cite{hcav} using a phase separation model:
defect cells contribute to the slow relaxing central peak
while coherent lattice motions (phonons) to the resolved 
inelastic channel. 
This mechanism has been applied with success to a wide class of 
disordered relaxor ferroelectrics\cite{Courtens82,Burns83}.

For La$_2$Cu$_{0.94}$Li$_{0.06}$O$_4$,
we envision that disorder accompanying doping
prevents the long-range order of the AF phase mainly by
upsetting interlayer magnetic phase coherence,
see Fig.~\ref{fig3} for the {\bf q}-distribution of
frozen spins. This upsetting is not uniform in the Griffiths
fashion\cite{qsg_hy} with weak and strong coupling parts in the 
sample. In our laminar material, however, the weak and strong coupling
parts have different dimensionality: 2D and nearly 3D, respectively.
The 2D part is a spin liquid and represents essentially the whole 
system at high temperature, see Fig.~\ref{fig4}. 
Part of sample with stronger interplane coupling tends to order 
three dimensionally below $\sim$150~K, producing
$S^{3D}({\bf q},E)$ and $S^{q3D}({\bf q},E)$. 
The condensation of the 2D spin liquid at $\sim$ 150~K 
into the 3D dynamic clusters
of diminishing energy scale, instead of a true long-range order,
may reflect the divergent fluctuations which destabilize
static order below the critical temperature
for random $XY$ or Heisenberg systems\cite{Hertz79}.
The nearly 3D spin-glass instead of a 3D antiferromagnet finally orders
at a much reduced $T_g\approx 20$~K, when $I^{q3D}+I^{3D}$
approaches the 2D spectral weight (Fig.~\ref{fig4}).
The phase separation into spin liquid and spin glass
components may be a consequence of no true mobility edge 
separating finite and infinite range correlations at 2D\cite{Hertz79}.
It would be interesting to extend the nonrandom 
theories\cite{2dheis} to incorporate
this disorder scenario and to explain 
the observed crossover from the $E/T$ to
a constant $E$ scaling in the 2D spin 
liquid\cite{bao02c,bao04a}. 

In summary, spins in La$_2$Cu$_{0.94}$Li$_{0.06}$O$_4$
develop {\it dynamic} AF order in the CuO$_2$ plane 
with very long $\xi_{\Box}$ below 150~K. The characteristic
energy of the 2D spin fluctuations is 0.18$k_B T$ for $T>50$~K and
1~meV for $T<50$~K\cite{bao02c}. Below $\sim$150~K,
interlayer phase coherence appears between some of these planar
AF clusters with an energy scale smaller than 70~$\mu$\,eV.
While the 2D AF correlations in an individual plane remain
liquid down to 1.5~K, coherent multiplane
AF correlations become glassy below $T_g$.
The phase separation into 2D spin-liquid and ``3D'' spin-glass
is most likely related to quasi-2D nature of magnetic exchange in 
the cuprates and is distinctly different from conventional spin-glasses. 
The theory of the spin-glass phase in doped cuprates should consider 
interlayer coupling, despite it is weak. 
Investigation of disorder effect on quantum
critical spin fluctuations is called for. The
phase separation, instead of a uniform magnetic phase,
suggests a possibility that superconductivity and the almost
3D AF order reside in different phases in La$_{2-x}$Sr$_x$CuO$_4$ 
and Y$_{1-x}$Ca$_x$Ba$_2$Cu$_3$O$_{6+y}$.

We thank R.H.\ Heffner, S.M.\ Shapiro and L.\ Yu for useful discussions. 
SPINS and NG7-SANS are supported partially by NSF. Work at LANL is
supported by DOE.


\end{document}